\documentclass[useAMS,usenatbib]{mn2e}
\usepackage{graphicx}
\usepackage{epsfig}
\usepackage{times}
\usepackage{natbib}
\usepackage{amsfonts}

\newif\ifAMStwofonts
\def\minf{$\dot{m}$}
\def\mco{$\dot{m}_{\rm co}$}

\def\m0mc{$\dot{m}/\dot{m}_{\rm c}$}

\def\rin{$r_{\rm in}$}

\title{Episodic Accretion on to Strongly Magnetic Stars}
\author[C. R. D'Angelo \& H. C. Spruit] {Caroline R. ~D'Angelo$^1$ and
  Hendrik C. ~Spruit$^1$ \\ $^1$ Max Planck Institute for
  Astrophysics, Garching, Germany} \date{Accepted 20.  Received
   ; in original form}

\pagerange{\pageref{firstpage}--\pageref{lastpage}}
\pubyear{2010}

\begin{document}

\maketitle

\label{firstpage}

\begin{abstract}
Some accreting neutron stars and young stars show unexplained episodic
flares in the form of quasi-periodic oscillations or recurrent
outbursts. In a series of two papers we present new work on an
instability that can lead to episodic outbursts when the accretion
disc is truncated by the star's strong magnetic field close to the
corotation radius (where the Keplerian frequency matches the star's
rotational frequency). In this paper we outline the physics of the
instability and use a simple parameterization of the disc-field
interaction to explore the instability numerically, which we show can
lead to repeated bursts of accretion as well as steady-state
solutions, as first suggested by Sunyaev and Shakura. The cycle
time of these bursts increases with decreasing accretion rate. These
solutions show that the usually assumed `propeller' state, in which
mass is ejected from the system, need not occur even at very
low accretion rates.
\end{abstract}


\section{Introduction}

The interaction between a strong stellar magnetic field and an
accretion disc can affect both the evolution and observational
properties of the star. Close to the star the field is strong enough
that the accretion disc is truncated, and mass is channelled along
field lines to accrete on to the star's surface. At the inner edge of
the truncated disc, the field and disc interact directly over some
finite region, allowing angular momentum exchange from the
differential rotation between the Keplerian accretion disc and the
star.

Angular momentum exchange between the field and the disc leads to two
different states that can exist for a disc truncated by a magnetic
field. The distinction depends on the position of the truncation
radius relative to the corotation radius, $r_{\rm c} \equiv
(GM_*/\Omega^2_*)^{1/3}$ (where $M_*$ and $\Omega_*$ are respectively
the mass and spin frequency of the star), the radius at which the
Keplerian frequency in the disc equals the star's rotational
frequency. If the disc is truncated inside $r_{\rm c}$ then the field-disc
interaction extracts angular momentum from the disc and accretion can
proceed. If on the other hand the disc is truncated outside $r_{\rm c}$, the
star-field interaction will create a centrifugal barrier that inhibits
accretion. This is usually called the `propeller regime', under the
assumption that most of the mass in the disc is expelled as an
outflow \citep{1975A&A....39..185I}.

Accreting stars with strong magnetic fields such as T Tauri stars,
and X-ray millisecond pulsars show a large degree of
variability in luminosity (corresponding to changes in accretion
rate), which may be ascribable to magnetic activity. For example, the
protostar EX Lupi (the prototype of the `EXor' class), a TTauri star,
increases and decreases in brightness by several magnitudes every 2--3
years \citep{2007AJ....133.2679H}. At much higher energies, a 1 Hz
quasi-periodic oscillation (QPO) in accreting millisecond pulsar SAX
J1808.8-3658 has been observed during the decay phase of several
outbursts \citep{2009ApJ...707.1296P}. The time-scale and magnitude of
the variability in both sources suggest changes in accretion rate in the inner regions
of the accretion disc, where it interacts with the star's magnetic
field.

In this paper we revisit a disc instability first suggested in
\cite{1977PAZh....3..262S} and developed in
\cite{1993ApJ...402..593S} (hereafter ST93), which can lead to
episodic bursts of accretion. The instability arises when the magnetic
field truncates the disc near the corotation radius. The magnetic
field initially truncates the disc outside but close to the corotation
radius, thus transferring angular momentum from the star to the disc
and inhibiting gas from accreting on to the star (the propeller
state).  However, close to $r_{\rm c}$, the energy and angular
momentum transferred by the field to the gas will not be enough to
unbind much of the disc mass from the system and drive an
outflow. Instead, the interaction with the magnetic field will prevent
accretion \citep{1977PAZh....3..262S}. As gas in the inner regions of
the disc piles up, the local gas pressure increases, forcing the inner
edge of the disc to move inwards until it crosses $r_{\rm c}$. When
the inner region of the disc cross inside $r_{\rm c}$, the centrifugal
barrier preventing accretion disappears (since now the differential
rotation between star and disc has changed sign) and the accumulated
reservoir of gas is accreted on to the star.  Once the reservoir has
been accreted, the accretion rate through the disc's inner edge
decreases, and the disc will again move outside $r_{\rm c}$, allowing
another cycle to start.

We study this process by following the time evolution of a thin
axisymmetric viscous disc, with a paramaterization of the interaction
between the disc and the magnetic field both inside and outside
$r_{\rm c}$. This approach allows us to investigate the behaviour of the
disc on time-scales much longer than the rotation period of the
star. Long time-scales are important since the instability evolves on
viscous rather than dynamical time-scales of the disc. We are able to
reduce the uncertainties in the detailed MHD interaction between the
field and the disc to two free (but constrained) parameters. Using
this description we can then investigate the physical conditions for
which the instability develops.

In this paper we describe in detail the physics that can lead to
episodic bursts of accretion and give a brief overview of the observed
oscillations. In a later paper we will explore the range of outbursts
seen in our simulations in more detail, and discuss their prospects
for observability in specific stellar systems.

\section{Magnetosphere-Disc Interactions}

\subsection{Interaction region between a disc and magnetic field}
\label{sec:global}

\begin{figure}
\includegraphics[width=\hsize]{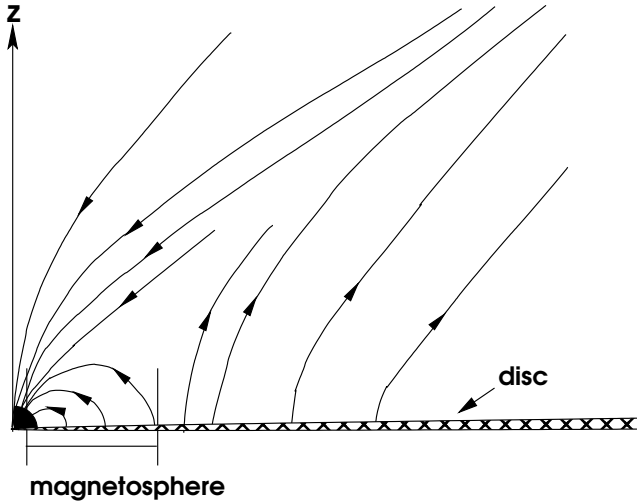}
\caption{Global magnetic field configuration for a strongly magnetic
  star surrounded by an accretion disc. In this picture, the majority
  of the field exists in an open configuration, and the connected
  region between the field and the disc is very small. Adapted from
  Lovelace et al. (1995).\label{fig:field}}
\end{figure}

We consider a star with a strong dipolar magnetic field surrounded by
a thin Keplerian accretion disc. We assume that the dipole is aligned
with both the star's spin axis and the spin axis of the disc, so that
the system is axisymmetric. Near the surface of the star the magnetic
field will truncate the disc, forcing gas into corotation with the
star. This inner region (in which the gas dynamics is regulated by the
magnetic field) is called the magnetosphere, and we define the {\em
  magnetospheric radius}, $r_{\rm m}$ as the radius at which the magnetic
field is no longer strong enough to force the disc into corotation
\citep{1993ApJ...402..593S}. Outside $r_{\rm m}$ the magnetic field will
penetrate the disc and become strongly coupled over some radial
extent, which we call the {\em interaction region}, $\Delta r$. Beyond
the interaction region the disc and magnetic field are decoupled, so
that the outer parts of the disc are not directly affected by the
stellar magnetic field. Figure \ref{fig:field} shows a schematic
picture for the magnetic field configuration, with a closed
magnetosphere close to the star, and a large region of opened field
lines further out.

In the interaction region, the differential rotation between the
Keplerian disc and star shears the magnetic field, generating an
azimuthal component $B_\phi$ from the initially poloidal field. This
in turn creates a magnetic stress which exerts a torque on the disc,
transferring angular momentum between the disc and star. The torque
per unit area exerted by the field on the disc is given by
${\rm}d{\mathbf tau}/{\rm d}r = rS_{z\phi}{\bf \hat{z}}$, where

\begin{equation}
\label{eq:stress}
S_{z\phi} \equiv \pm \frac{B_\phi B_z}{4\pi}
\end{equation}
is the magnetic stress generated by the twisted field lines. The sign
of the torque will depend on the location of the coupled disc region
relative to the corotation radius, $r_{\rm c} \equiv
(GM_*/\Omega_*^2)^{1/3}$. If the coupling takes place inside $r_{\rm
  c}$ the torque will extract angular momentum from the disc, spinning
down the disc (and spinning up the star), while if the coupling is
outside $r_{\rm c}$ the torque adds angular momentum to the disc,
spinning it up (and spinning down the star).

The radial extent of the interaction region has been a point of
long-standing controversy in the study of accretion discs. In an early
series of influential papers, Ghosh et al. (1977;
\citealt{1979ApJ...232..259G,1979ApJ...234..296G}) argued that the
coupled region is large ($\Delta r / r \gg 1$), so that the magnetic
field exerts a torque over a considerable fraction of the disc with a
resulting large influence on the spin evolution of the star. However,
the original model proposed by Ghosh \& Lamb was shown to be
inconsistent by \cite{1987A&A...183..257W}, since the magnetic
pressure they derived from field winding far from $r_{\rm c}$~is high
enough to completely disrupt the majority of the disc.

More recent analytical and numerical work has shown that the
interaction region is likely much smaller, and much of the disc is
disconnected from the star (see \citealt{2004Ap&SS.292..573U} for a
recent review). This comes about from the fact that in force-free
regions (where the magnetic pressure dominates over the gas pressure)
as are likely to exist above an accretion disc, field lines will tend
to open up as the twisting increases \citep{1985A&A...143...19A,
  1994MNRAS.267..146L}. As the disc and star rotate differentially,
the increasing twist $\Delta \phi$ in the field line will only
increase the $B_\phi$ component to some maximum $B_\phi \sim B_z$
before the increased magnetic pressure above the disc causes the field
lines to become inflated and eventually open, severing the connection
between the disc and star. Analytic studies of a sheared force-free
magnetic field \citep{1985A&A...143...19A, 1994SSRv...68..299V,
  2002ApJ...565.1191U} have shown that the $B_\phi$ component will
grow to a maximum twist angle $\Delta \phi \sim \pi$ before
opening. The twist angle grows on the time-scale of the beat frequency
$\equiv |\Omega_*-\Omega_K|^{-1}$, which is very short compared to the
viscous time-scale in the disc except in a very small region around
corotation.

To prevent field lines from opening, they must be able to slip through
the disc faster than the rate at which the field is being wound up.
The rate at which the field can move through the disc is set by the
effective diffusivity, $\eta$, of the disc. Like the effective
viscosity, $\nu$, that drives the transport of angular momentum, the
effective diffusivity is also assumed to be driven by turbulent
processes in the disc. Recent numerical studies of MRI
(Magnetorotational Instability) turbulence (believed to be responsible
for angular momentum transport in at least the inner regions of
accretion discs) have tried to measure $\eta$ directly. In these
simulations, an external magnetic field is imposed on a shearing box
simulation, and the effective magnetic diffusivity is estimated as the
flow becomes unstable. The results suggest that the effective
diffusivity and viscosity are of similar size, that is, the effective
magnetic Prandtl number, $Pr \equiv \nu/\eta$ is of order unity
\citep{2009A&A...507...19F}. Such a large magnetic Prandtl number
implies that for realistic disc parameters the magnetic field will not
be able to slip through the majority of the disc fast enough enough to
prevent field lines from opening \citep{1995MNRAS.275..244L,
  2002ApJ...565.1191U}. Outside this region there will still be some
coupling between the disc and the star as the gas moves from
Keplerian to corotating orbits, but this estimate suggests that the
actual extent of coupling is small ($\Delta r/r < 1$) regardless of
where the disc is truncated relative to the corotation radius.

Once the field lines are opened, there may be some reconnection across
the region above the disc between open magnetic field lines
(e.g. \citealt{1990A&A...227..473A,1997ApJ...489..199G,2002ApJ...565.1191U}). The
effective size of the interaction region would then depend on the
efficiency of reconnection, and could also then become time-dependent
(although likely on time-scales of order the dynamical time, which is
much shorter than the viscous evolution time-scale). The opening and
reconnection of field lines has also been suggested as a possible
launching mechanism for strong disc winds and a jet
(e.g. \citealt{1990A&A...227..473A,1996ApJ...468L..37H,1997ApJ...489..199G}). This
picture of a small interaction region with some reconnection was first
proposed by \cite{1995MNRAS.275..244L}, and has been supported by 2
and 3D simulations of accretion discs interacting with a magnetic
field
(e.g. \citealt{1997ApJ...489..890M,1997ApJ...489..199G,1996ApJ...468L..37H,2009arXiv0907.3394R}).

In summary, although the extent of the interaction region is uncertain
(subject to uncertainties in the effective diffusivity of magnetic
field in the disc and its possible reconnection in the magnetosphere,
as well as the detailed interaction between the disc and field near
the magnetosphere), numerical and analytic work suggests that it is
small. Except for very special geometries for the magnetic field (such
as \citealt{2000MNRAS.317..273A, 1994ApJ...429..781S}), the low
effective magnetic diffusivity in the disc will force the magnetic
field into a largely open configuration, and the majority of the
accretion disc will be decoupled from the star, in strong contrast to
the prediction of the \cite{1979ApJ...232..259G} model.

The extent of the interaction region as well as the average magnitude
of the $B_\phi$ component generated by the disc-field interaction will
depend on the detailed interaction between the disc and the field as
the gas moves from Keplerian orbits to corotation with the star, as well
as the frequency and magnitude of possible reconnection events. In the
present work we therefore assume that the time-averaged $B_\phi$
component generated by field-line twisting will be some constant
fraction of $B_z$, so that $B_\phi/B_z \equiv \eta < 1$. We also
assume that $\Delta r/r $ is small ($< 1$) but leave it as a free
parameter.

\subsection{Accretion and angular momentum transport}
\label{sec:ang_mom}
In this paper we describe the evolution of an accretion disc in which
the conditions at the inner boundary are changing in time. Before
doing this, however, we review how the conditions at the inner
boundary affect the angular momentum transport and density structure
of a thin accretion disc. In the thin-disc limit the evolution
equation for the surface density $\Sigma$ can be written:

\begin{equation}
\label{eq:sigma_ev}
\frac{\partial \Sigma}{\partial t} =
\frac{3}{r}\frac{\partial}{\partial r}[r^{1/2}\frac{\partial}{\partial
    r}(r^{1/2}\nu\Sigma)],
\end{equation}
where $\nu$ is the effective viscosity in the disc that enables
angular momentum transport. In a steady state (in which the accretion
rate is constant throughout the disc), the general solution for
$\nu\Sigma$ is given by:

\begin{equation}
\label{eq:steady}
\nu\Sigma = \frac{\dot{m}}{3\pi}\left(1-\beta\left(\frac{r_i}{r}\right)^{1/2}\right),
\end{equation}
where $r_i$ is the inner edge of the disc, \minf is the accretion rate
and $\beta$ is a dimensionless measure of the angular momentum flux
through the disc per unit mass accreted \citep{1991ApJ...370..604P,
  1991ApJ...370..597P}.

All accretion discs have a boundary layer at their inner edge that
connects the disc with either the surface of the star or the star's
magnetosphere. In the boundary layer the gas must transition from
Keplerian orbits to orbits corotating with the star in order to
accrete. The structure of this boundary layer will determine the value
of $\beta$ in (\ref{eq:steady}). In the standard accretion
scenario, that is, for accretion on to a slowly-rotating star or on to
the star's magnetosphere inside the corotation radius, the gas in the
boundary layer will be decelerated, meaning that there will be a
maximum in the rotation profile, $\Omega(r)$.  At the maximum in
$\Omega(r)$, there is no longer an outward transfer of angular
momentum from viscous torques, which in the thin-disc approximation
will cause the surface density to decrease sharply, so that $\beta$ =
1 in (\ref{eq:steady}) \citep{1972A&A....21....1P,
  1973A&A....24..337S}. The maximum in $\Omega(r)$ effectively
corresponds to the inner radius of the disc, since inside this radius
gas is viscously decoupled from the rest of disc. The gas falling
through the inner boundary of the disc will add its specific angular
momentum ($\dot{m}r^2_{\rm in}\Omega$) to the star, spinning it up.

\begin{figure}
\includegraphics[width=\hsize]{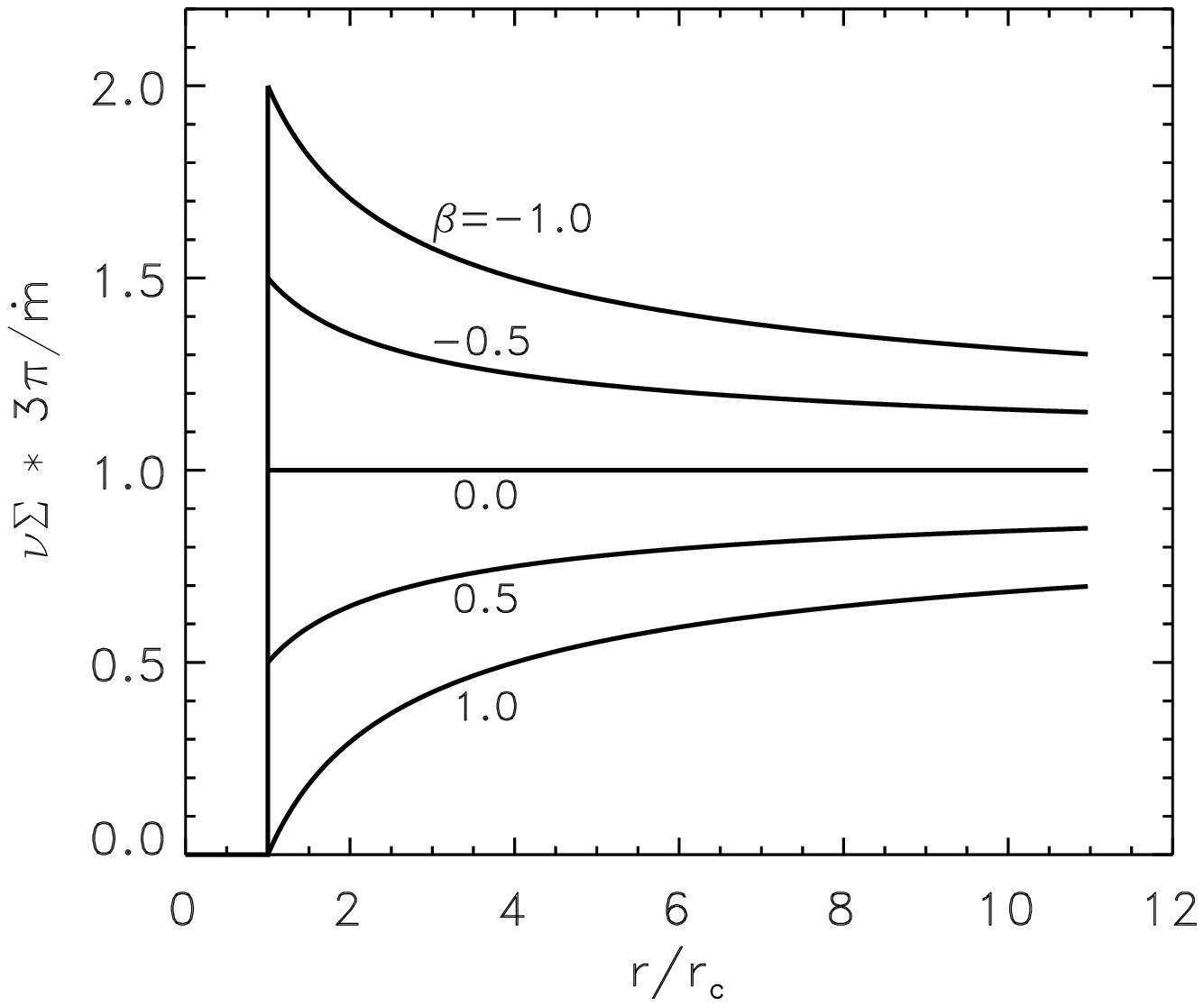}
\includegraphics[width=\hsize]{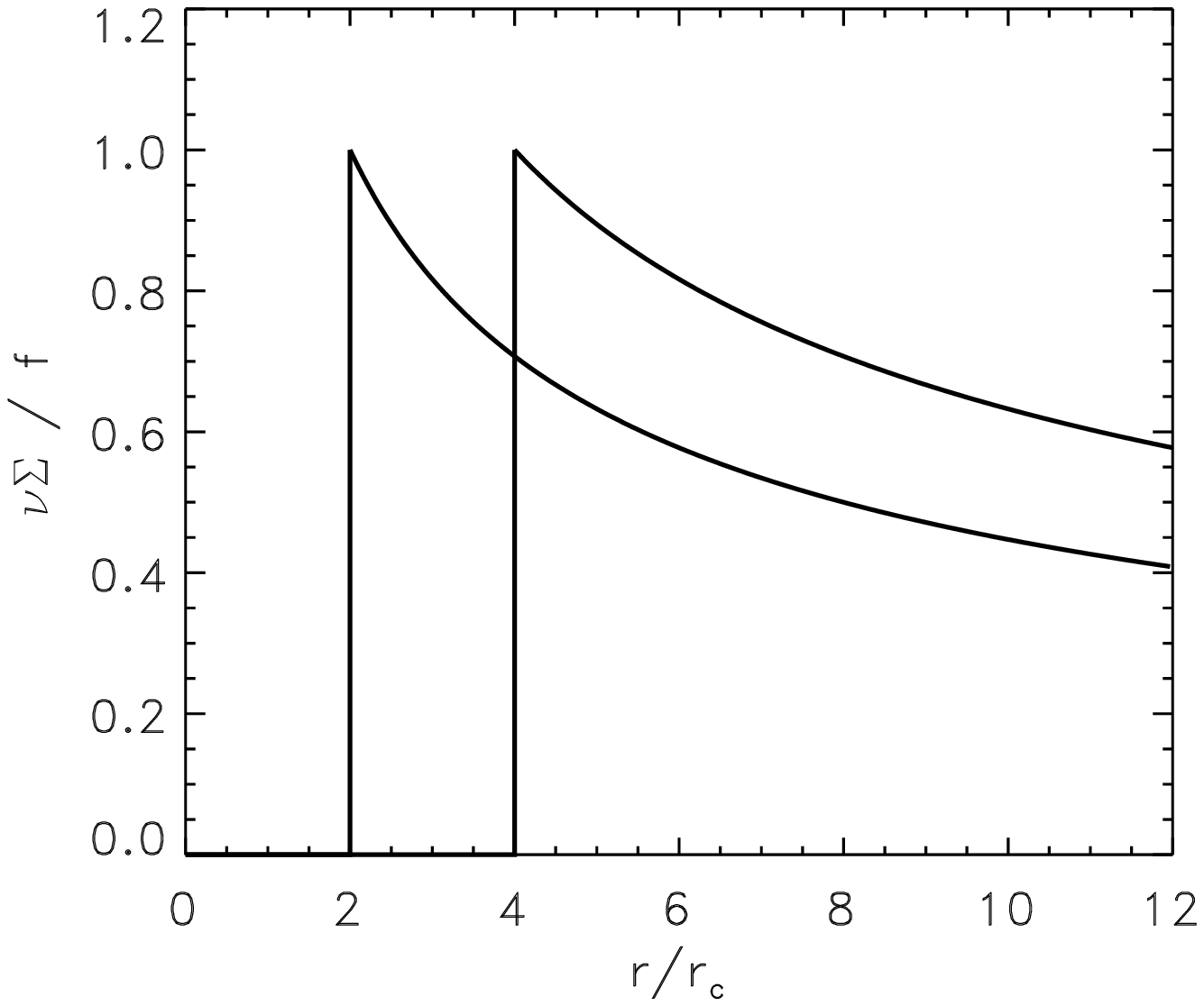}
\caption{Surface density $\nu\Sigma$ of a thin disc as a function of
  distance from the corotation radius $r_{\rm c}$, for a steady, thin
  viscous disc. Top: steady accretion at a fixed accretion rate $\dot
  m$, for inner edge of the disc at corotation. $\beta$ measures the
  angular momentum flux, $\beta= 1$ corresponding to the standard case
  of accertion on to a slowly rotation object. For $\beta < 0$ the
  angular mometum flux is outward (spindown of the star). Bottom:
  `quiescent disc' solutions with $\dot{m} = 0$ and a steady outward
  angular momentum flux due to a torque $f$ applied at the inner
  edge. The two curves show solutions for $ r_{\rm in}/r_{\rm c}
    = $ 2 and 4.\label{fig:sigma}}
\end{figure}

However, there are in fact a wide range of solutions for the surface
density profile of an accretion disc depending on the conditions
imposed by the boundary layer, which in turn set the rate of angular
momentum transport across the inner boundary of the disc. In a
nonmagnetic star spinning close to breakup
\citep{1991ApJ...370..597P,1991ApJ...370..604P}, the angular momentum
flux can be inward or outward, depending on the accretion history of
the star. The dimensionless angular momentum flux $\beta$ can in
principle have any value less than 1 in this case. The top panel of
Fig. \ref{fig:sigma} shows the steady-state surface density profile
for a range of different values of $\beta$ from -1 to 1.

\cite{1977PAZh....3..262S} studied a similar situation in which there
is outward angular momentum transport in an accretion disc, and showed
adding angular momentum at the inner edge of the accretion can in fact
halt accretion altogether. The evolution of the disc in this case
depends on the rate at which angular momentum is being injected at the
inner edge of the disc compared to the rate at which it is carried
outwards via viscous coupling. If angular momentum is injected into
the inner boundary of the disc at exactly the same rate as viscous
transport carries it outwards, then all accretion on to the star will
cease. For a steady state like this to exist, the outward angular
momentum flux due to the magnetic torque at the inner edge of the disc
has to be taken up at some larger distance. In a binary system, this
sink of angular momentum can be the orbit of the companion star. If
the disc is sufficiently large, the angular momentum can also be taken
up by the outer parts of the disc, while the inner parts of the disc
are close to a steady state. The inner edge of the disc then slowly
moves outward under the influence of the angular momentum flux. The
surface density distribution in this case can be found from
(\ref{eq:steady}) by taking the limit $\dot m\rightarrow 0$, while
letting $\beta\rightarrow -\infty$ (noting that it measures the
angular flux per unit accreted mass). This yields:

\begin{equation}
\label{eq:steady_sigma}
\nu\Sigma = f(r_i)\left(\frac{r_i}{r}\right)^{1/2},
\end{equation}
where $f(r_{\rm i})$ is a measure of the torque exterted at the
inner edge of the disc. The bottom panel of Fig. 2 shows the surface
density, scaled to the value of $f(r_{\rm i})$, for two instances
  of (\ref{eq:steady_sigma}) with different values of $r_{\rm in}$.

\cite{1977PAZh....3..262S} refer to this solution as a `dead disc',
since there is no accretion on to the star. In this paper we call
non-accreting discs without large outflows `quiescent discs', to avoid
confusion with `dead zones' thought to be present in proto-stellar
discs (regions in which there is insufficient ionization to drive
angular momentum transport via MRI but are too hot for efficient
angular momentum transport via gravitational instabilities;
e.g. \citealt{1996ApJ...457..355G}). These quiescent discs play a role
in the cyclic solutions discussed in Section \ref{sec:model}. In these solutions
accreting phases are separated by long intervals in which the inner
disc is close to the quiescent state described by (\ref{eq:steady_sigma}).

\subsection{Evolution of a disc truncated inside the corotation radius}
\label{sec:rin<rcorot}

When the accretion disc is truncated by a magnetic field inside the
corotation radius, the standard $\beta = 1$ case applies for a
steady-state solution. The location of the inner edge of the disc
\rin will be determined by the interaction between the disc and
magnetic field, and change with changing conditions at the inner edge
(such as the accretion rate on to the star). Here we estimate the
location of \rin, and use it to show how the inner boundary of the
disc will change in a non-steady disc.

We define the inner edge of the disc as the point at which material in
the disc is forced into corotation with the star. We use the
azimuthal equation of motion for gas at the magnetospheric radius to
obtain an estimate for \rin in a disc (see, e.g. ST93):

\begin{equation}
\label{eq:eq_mo}
2\pi\Sigma \frac{\partial}{\partial
  t}(rv_\phi)-\frac{\dot{m}_{\rm in}}{r}\frac{\partial}{\partial r}(rv_\phi) +
2\pi rS_{z\phi} = 0,
\end{equation} 
where $\dot{m}_{\rm in} = -2\pi r\Sigma v_r$ is the accretion rate through the
inner edge of the disc. (\ref{eq:eq_mo}) neglects viscous
angular momentum transport through the inner regions of the disc,
under the assumption that it will be much smaller than angular
momentum transport from the magnetic field. Using $v_\phi = \Omega_*
r$ (since at $r_{\rm in}$ the gas corotates with the star), and
assuming a steady-state solution ($\partial/\partial t = 0$), 
(\ref{eq:eq_mo}) becomes:

\begin{equation}
\label{eq:rm1}
\frac{\dot{m}\Omega_*}{\pi} = r_{\rm in}S_{z\phi} = \frac{r_{\rm in}B_\phi B_z}{4\pi},
\end{equation}
where $S_{z\phi}$ is the magnetic stress from the coupling between
the disc and star (introduced in Section \ref{sec:global}). As
long as the wind-up time for the field is shorter than the rate at
which \rin~is changing, $B_{\phi}/B_z$ will be roughly constant, so we
make the assumption that $B_\phi = \eta B_z$, where $\eta < 1$ and is
constant.

For a dipole field aligned with the star's axis of rotation ($B_z =
\mu/r^3$, where $\mu = B_SR^3_*$ is the star's magnetic dipole
moment), (\ref{eq:rm1}) can be re-written:

\begin{equation}
\label{eq:rm}
r_{\rm in} = \left(\frac{\eta\mu^2}{4\Omega_*\dot{m}_{\rm in}}\right)^{1/5}.
\end{equation}
For $\eta = 0.1$, this estimate gives a value for $r_{\rm in}$ about 40\%
smaller than the simple estimate found by equating the magnetic
pressure from the field ($B^2/8\pi$) to the ram pressure from
spherically-symmetric gas in free-fall on to the star (e.g.
\cite{1972A&A....21....1P}).

The derivation for \rin~above holds for steady accretion. For the
problem studied here the position of the inner edge (set by the
  location of the magnetosphere) will change in time, which requires
a minor reinterpretation of (\ref{eq:rm}). If \rin~is moving in time,
the mass flux $\dot{m}_{\rm co}$ in the reference frame comoving with
\rin~differs from the mass flux, \minf, measured in a fixed frame:

\begin{equation}
\label{eq:dotm}
\dot{m}_{\rm co} = \dot{m} + 2\pi r\Sigma \dot{r}_{\rm in},
\end{equation}
where $\dot{r}_{\rm in}$ is the time derivative of \rin. 

Since the torque between the magnetosphere and the disc acts at the
inner edge, the mass flux entering the magnetosphere (used in
(\ref{eq:rm})) is given by $\dot{m}_{\rm co}$, not \minf. As before,
\minf~itself is given in terms of the surface density by the usual
thin disc expression:

\begin{equation}
\dot{m} = 3r^{1/2}_{\rm in}\frac{\partial}{\partial r}\left(r^{1/2}\nu\Sigma\right)\big|_{r_{\rm in}}.
\end{equation}

\subsection{Evolution of a disc truncated outside the corotation radius} 
\label{sec:rin>rcorot}

If the star is spinning fast enough, the magnetic field can truncate
the disc {\em outside} $r_{\rm c}$. In this case the interaction with
the magnetic field will {\em add} angular momentum to the disc,
creating a centrifugal barrier that inhibits accretion. This scenario
was first described by \cite{1975A&A....39..185I} and is often termed
the `propeller' regime, under the assumption that the interaction
with the magnetic field will expel the disc at \rin~as an
outflow via the `magnetic slingshot' mechanism
\citep{1982MNRAS.199..883B}. 

However, in order for the gas to be ejected from the system, it must
be accelerated to at least the escape speed ($v_{esc} =
\sqrt{GM_*/2r}$). At the inner edge of the interaction region the gas
is brought into corotation with the star, where $v_{\rm c} =
\Omega_*r$. If this is less than the escape speed, the majority of the
gas will not be accelerated enough to be expelled. Setting $v_{esc} =
v_{\rm c} = \sqrt{GM_*/r^3_{\rm c}}r$ implies that for \rin $<
1.26 r_{\rm c}$ most of the gas will not be expelled. 

Part of the disc could still be expelled in an outflow, but while the
majority of the gas remains confined in the disc, the disc can act as
an efficient sink for angular momentum from the star and accretion can
effectively be halted. The open field lines at larger radii could
launch a disc wind which would provide an additional sink for angular
momentum and somewhat change the structure of the disc (e.g.
\citealt{2005ApJ...632L.135M}). Numerical studies of the field-disc
interaction, for example, find that reconnection across field lines
can lead to intermittent accretion
(e.g. \citealt{1997ApJ...489..199G}, see also Section
\ref{sec:discussion}). However, models of disc winds typically include
mass loss rate as a parameter of the problem, so that the amount of
mass actually lost to the wind is uncertain.  In this paper we make
the assumption that the disc becomes quiescent, that is, for $r_{\rm
  in}> r_{\rm c}$ no accretion or outflows occur. The steady-state
disc solution is then given by (\ref{eq:steady_sigma}).

In the next section we will derive $f(r_{\rm in})$, the boundary
condition for the surface density at the inner edge of a
quiescent disc. Like for cases when $r_{\rm in} < r_{\rm c}$, we
want to study non-steady-state solutions in which \rin~moves in
time. As in the steady-state case, to derive $\dot{r}_{\rm in}$
we consider the difference in accretion rate at \rin~in a fixed frame
and in a frame comoving with \rin. Since for a quiescent disc no
matter is being accreted on to the star, \mco = 0, so that
(\ref{eq:dotm}) can be written:

\begin{equation}
2\pi r\Sigma\dot{r}_{\rm in} = -3r^{1/2}_{\rm in}\frac{\partial}{\partial
  r}\left(r^{1/2}\nu\Sigma\right)\big|_{r_{\rm in}}.
\end{equation}
Together with (\ref{eq:sigma_ev}), a viscosity prescription and
condition for the outer boundary, we can use the results from this
section and the previous one to study the time-dependent behaviour of
an accretion disc interacting with a magnetic field.

\section{Cyclic accretion}
\label{sec:model}

The existence of quiescent disc solutions can naturally lead to
bursts of accretion. Since there is very little accretion on to the
star or outflow, if mass continues to accrete from larger radii it
will pile up in the inner regions in the disc until the gas pressure
is high enough to overcome the centrifugal barrier from the magnetic
field-disc interaction and accretion can proceed. Once the reservoir
has been emptied the inner edge of the disc will move back outside the
corotation radius and the reservoir will start to build up again.

In Sections \ref{sec:rin<rcorot} and \ref{sec:rin>rcorot} we showed
how the inner radius of a thin viscous accretion disc will evolve
inside and outside corotation. To study the time-dependent evolution
of a disc, we must connect these two states as the inner edge of the
disc passes through the corotation radius. We also require a
description for $f(r_{\rm in})$, the inner boundary condition for the
disc truncated outside $r_{\rm c}$.

\subsection{Surface density profile for $r_{\rm in} > r_{\rm c}$}
\label{sec:rin_outsiderc}

When the interaction region is outside $r_{\rm c}$, the star is rotating
faster than the Keplerian disc and the magnetic field lines lead the
disc, adding angular momentum to the material in the inner regions. As
discussed in Section \ref{sec:global}, the torque per unit area
exerted on the disc will be $\langle S_{\phi z}\rangle r$, so that the
torque exerted across the entire interaction region (assuming it is
small) is approximately:

\begin{equation}
{\mathbf \tau} \simeq 4 \pi \langle S_{\phi z}\rangle r_{\rm in} \Delta r {\bf \hat{z}},
\end{equation}
where the extra factor 2 comes from coupling to both sides of the
disc.

As argued in the previous section, if the disc is truncated close to
but outside $r_{\rm c}$, the majority of the gas in the interaction
region will not be expelled in an outflow. Instead, the angular momentum
from the magnetic field is transferred outwards to the rest of the
disc. We can derive a relationship between the position of and surface
density at the inner edge of non-interacting disc from the
conservation of angular momentum across the interaction region.

Since the interaction region is small we do not consider its density
profile explicitly, focusing instead on its influence on the
non-interacting disc. We therefore define $r_{\rm in}$ as the point in
the disc just outside the interaction region, where there is no
magnetic coupling between the disc and the star. Across the
interaction region the density in the disc decreases sharply (since
the gas is forced into nearly corotating orbits with the star).  We
make the simplifying assumption that none of the mass in the disc
escapes, either into an outflow or through the magnetosphere on to the
star. The inner edge of the interaction region, $r_{\rm in} - \Delta
r$, is therefore defined as the point at which the surface density
drops to zero.

To determine $\Sigma$ at \rin we consider the angular momentum flux
across $\Delta r$ when $r_{\rm in} > r_{\rm c} $. The flux of angular momentum must
be continuous across $\Delta r$, meaning that the viscous angular
momentum transport outside $\Delta r$ must balance the angular
momentum flux added by the magnetic field across the interaction
region.  This balance is written:

\begin{eqnarray}
\label{eq:ang_mom}
\lefteqn{\dot{m} r^2\Omega - 2\pi r(\nu\Sigma)^+r^2\Omega'
  =}\\ \nonumber&& \dot{m} r^2\Omega - 2\pi r(\nu\Sigma)^-r^2\Omega'
+ \int^{r_{\rm m}+\Delta r}_{r_{\rm m}} 4 \pi r^2S_{z\phi} dr.
\end{eqnarray}

In this equation, $\nu^{\pm}$ and $\Sigma^{\pm}$ are the viscosity and
surface density inside ($^-$) and outside ($^+$) $\Delta r$, $\dot{m}
= 2\pi r(\Sigma v_r)^{\pm}$ is the mass flux through $\Delta r$~
(where $v_r$ is the radial velocity of the gas) and $\Omega$ is the
orbital frequency at \rin. The first term on either side of the
equation denotes the advection of angular momentum across \rin, while
the second is the angular momentum transported by viscous
stresses. The final term on the right hand side is the angular
momentum added by the magnetic field to the coupled region of the
disc.  The first term on both sides cancel (to enforce conservation of
mass across $\Delta r$), and we make the further assumption that in
the interaction region most of the angular momentum is transported
through external magnetic torques rather than viscous stress, so that
$(\nu\Sigma)^- \ll (\nu\Sigma)^+$. For a small interaction region, the
last term in (\ref{eq:ang_mom}) can be re-written:

\begin{equation}
\label{eq:ang_mom2}
\int^{r_{\rm in}}_{r_{\rm in} - \Delta r} 4 \pi rS_{z\phi} dr \approx 4\pi \Delta r
r_{\rm in}\langle S_{z\phi}\rangle.
\end{equation}

(\ref{eq:ang_mom}) can then be re-written to yield the surface
density at \rin~for $r > r_{\rm c}$:

\begin{equation}
\label{eq:sigma_in}
(\nu\Sigma)^+ = -\frac{2 \langle S_{z \phi}\rangle \Delta r}{\pi
  r_{\rm in}\Omega'}.
\end{equation}

As predicted in Section \ref{sec:rin>rcorot}, (\ref{eq:sigma_in})
shows that the surface density at \rin~will be large, a consequence of
the torque being applied by the disc-magnetic field coupling
\citep{1977PAZh....3..262S, 1991ApJ...370..604P,
  1991ApJ...370..597P}. (\ref{eq:sigma_in}) corresponds to the
function $f(r_{\rm in})$ introduced in Section \ref{sec:ang_mom} for
$r_{\rm in} > r_{\rm c}$, that is, the boundary condition at the
  inner edge of the disc. In a time-dependent system, as gas accretes
from larger radii (via viscous torques) it will pile up near \rin
and the increased gas pressure will push the inner edge of the disc
further inwards towards $r_{\rm c}$.

\subsection{Transition region}
\label{sec:transition}

When the inner edge $r_{\rm in}$ is well inside $r_{\rm c}$,
conditions at the inner edge are the standard ones for accretion of a
thin disc on a slowly rotating object:

\begin{equation}
 \Sigma(r_{\rm in})=0, 
\end{equation}
while the time-dependent position of the inner edge is determined by
(\ref{eq:rm}):

\begin{equation}
\label{eq:mdot_in}
r_{\rm in} = \left(\frac{\eta \mu^2}{4\Omega_*\dot{m}_{\rm co}}\right)^{1/5},
\end{equation}
where $\dot m_{\rm co}$ is the mass flux in a frame comoving with
$r_{\rm in}$ as discussed above. 

When the inner edge is outside the corotation radius, the
magnetosphere does not accrete: 
\begin{equation}
\dot m_{\rm co}=0,
\end{equation}
while the surface density at $r_{\rm in}$ is determined by a magnetic
torque, as discussed above. With the Keplerian value for
$\Omega(r_{\rm in})$ and assuming a dipolar magnetic field,
the results of Section \ref{sec:rin_outsiderc} can be re-written:

\begin{equation}
\label{eq:sigma0}
(\nu\Sigma)^+ = \frac{\eta\mu^2}{3\pi(GM_*)^{1/2}}\frac{\Delta
    r}{r^{9/2}_{\rm in}}.
\end{equation}

To connect these two limiting cases, we assume that the effect of the
interaction processes is equivalent to a smooth transition in the
conditions. This is valid since the time-scales we are interested in
are much longer than the orbital time-scale on which the conditions of
the transition region between disc and magnetosphere vary. The
assumption is thus that the effect of the fast processes in the
transition region can be represented by averages. The mass flux on to
the magnetosphere is therefore taken to vary smoothly from 0 for
$r_{\rm in}$ well outside corotation to the value in
(\ref{eq:mdot_in}) valid well inside:

\begin{equation}
\label{eq:mdot_co}
\dot {m}_{\rm co}= y_m \dot m^+,
\end{equation}
where $ \dot {m}^+$ is given by (\ref{eq:mdot_in}). For the
connecting function $y_m$ we take a simple function that varies from 0
to 1 across the transition: 

\begin{equation}
y_m = \frac{1}{2}\left[1 - \tanh\left(\frac{r_{\rm in}-r_{\rm c}}{\Delta r_2}\right)\right]
\end{equation}
where $\Delta r_2$ is the nominal width of the disc-magnetosphere
transition and a parameter of the problem.

Similarly the surface density at the inner edge makes a smooth
transition from its value in (\ref{eq:sigma0}) to 0:

\begin{equation}
\label{eq:sigma_in2}
 \Sigma_{\rm in}= y_\Sigma \Sigma^+,
\end{equation}
 where the connecting function $y_\Sigma$ is:

\begin{equation}
y_\Sigma = \frac{1}{2}\left[1 + \tanh\left(\frac{r_{\rm in}-r_{\rm c}}{\Delta r}\right)\right].
\end{equation}

All the uncertainties in the transition region are thus subsumed in
the parameters $\Delta r$ and $\Delta r_2$. In Section
\ref{sec:results} we study the effect of these uncertainties with a
parameter survey. The effective widths of the transition of
magnetospheric accretion rate and inner-edge surface density need not
be the same, and we in fact find that the difference between $\Delta
r$ and $\Delta r_2$ is important for the form of the resulting
accretion cycles.

\subsection{Physical constraints on $\Delta r$ and $\Delta r_2$}
\label{sec:constraints}

In this paper we treat $\Delta r$ and $\Delta r_2$ as free
parameters. However, a lower limit on both parameters
can be set by considering the stability of the inner regions of the
disc to the interchange instability. In the quiescent disc, the
low-density magnetosphere must support the high-density disc against
infall. This configuration will be unstable to interchange instability
(the analog of the Kelvin-Helmholtz instability), unless the surface
density gradient in the interaction region is shallow enough to
suppress it. This sets a limit on the minimum width of the interaction
region, $\Delta r$, where the density gradient falls from its maximum
(at \rin) to close to zero in the magnetosphere. 

This instability also sets a limit on the minimum width of $\Delta
r_2$, the transition length over which the disc moves from a
non-accreting quiescent disc to one in which there is accretion
through the inner boundary. As $r_{\rm in}$ moves closer to $r_{\rm
  c}$ the width of the interaction region preventing accretion
(i.e. where the field lines are adding angular momentum to the disc)
decreases. When the width of the interaction region outside $r_{\rm
  c}$ becomes smaller than is stable against the interchange
instability, accretion through the magnetosphere will begin.  $\Delta
r_2$ must therefore be larger or equal to this value, that is, at this
minimum distance from $r_{\rm c}$ accretion onto the star will take place.

\cite{1995MNRAS.275.1223S} studied the stability of a disc interacting
with a magnetic field to interchange instabilities, and derived the
following linear stability criterion:

\begin{equation}
\frac{B_rB_z}{2\pi\Sigma}\frac{\rm{d}}{\rm{d}r}\ln\left|\frac{\Sigma}{B_z}\right|
> 2\left(r\frac{\rm{d}\Omega}{\rm{d}r}\right)^2.
\end{equation}

Assuming that $B_r \sim B_\phi$, in our formulation this inequality
becomes:

\begin{equation}
\frac{3\alpha}{1+\tanh\left(\frac{\Delta r_2}{\Delta
    r}\right)}\left(\frac{H}{r}\right)^2 >
2\left(1-\left(\frac{r_{\rm in}}{r_{\rm c}}\right)^{3/2}\right)^2.
\end{equation}

For $\alpha = 0.1$ and assuming $H/r$ is in the range 0.07--0.1, the
range of $\Delta r/r = [0.05,0.1]$ will satisfy this inequality for
$\Delta r_2/r = [0.01,0.02]$. In this inequality larger values of
$\Delta r$ correspond to smaller possible values for $\Delta r_2$,
since larger $\Delta r$ correspond to smaller maximum $\Sigma(r_{\rm
  in})$ and hence shallower gradients. This instability has recently
been studied using 3D numerical simulations
\citep{2008MNRAS.386..673K}, who find numerically approximately the same criterion
for stability as \cite{1995MNRAS.275.1223S}. The shaded regions of Figs. \ref{fig:parmap1} and
\ref{fig:parmap2} show the values for $\Delta r_2$ and $\Delta r$ that
are unstable to the instability studied in this paper. The simple
analysis of this section suggests that at least part of the shaded
sections in Figs. \ref{fig:parmap1} and \ref{fig:parmap2} will be
stable against the interchange instability, so that the larger
magnetosphere-disc instability could occur.

\section{Numerical Implementation}
\label{sec:computation}

\subsection{Disc equation and viscosity prescription}
\label{sec:disc_eq}

To study the surface density evolution of an accretion disc
interacting with a magnetic field as outlined in the previous section,
we use a time-dependent numerical simulation of a diffusive accretion
disc. Our assumption that the interaction region is small ($\Delta r/r
< 1$) means that rather than calculate the disc behaviour in the
interaction region explicitly we can instead use the physics of the
interaction region to derive boundary conditions for the inner edge of
the non-interacting disc.

We assume that the accretion disc (outside the interaction region) can
be treated in the thin-disc limit, so that the evolution equation for
the surface density $\Sigma$ is given by (\ref{eq:sigma_ev}). We
assume that the viscosity in the disc follows a power-law dependence,
so that:

\begin{equation}
\label{eq:visc}
\nu = \nu_0 r^\gamma,
\end{equation}
where $\nu_0 = \alpha (GM_*)^{1/2}(H/R)^2$ and $\gamma = 0.5$
following the standard $\alpha$-viscosity prescription
\citep{1973A&A....24..337S}. To evolve (\ref{eq:sigma_ev}) in time, we
require boundary conditions at \rin~and $r_{\rm out}$, plus an
additional equation to describe the movement of the inner edge of the
disc, $\dot{r}_{\rm in}$. We set the outer boundary by defining the
mass accretion rate through the outer edge of the disc (\minf), which
we vary as a parameter of the problem.  This defines the
  time-averaged mass accretion rate in the disc. The surface density
at the inner edge of the disc is given by (\ref{eq:sigma_in2}):

\begin{equation}
\label{eq:sigma_in_final}
\Sigma(r_{\rm in}) =
\frac{\eta\mu^2}{6\pi(GM_*)^{1/2}\nu_0}\frac{\Delta
  r}{r^{9/2+\gamma}_{\rm in}}\left[\tanh\left(\frac{r_{\rm in}-r_{\rm c}}{\Delta r}\right) +
  1\right].
\end{equation}

We calculate the displacement of the inner boundary using the
results of Sections \ref{sec:rin<rcorot} and \ref{sec:rin>rcorot}, by
considering the difference between the total mass flux at \rin~in a
fixed and comoving frame of reference:

\begin{equation}
\dot{m}_{\rm co} = \dot{m} + 2\pi r \Sigma \dot{r}_{\rm in},
\end{equation}
where $\dot{m}_{\rm co}$ is given by (\ref{eq:mdot_co}). This
expression can be re-written: 
\begin{eqnarray}
\label{eq:mdot,final}
6\pi r^{1/2}_{\rm in}\frac{\partial}{\partial t}(\nu\Sigma r_{\rm in}) = -2\pi
r_{\rm in}\Sigma(r_{\rm in})\dot{r}_{\rm in} + \\ \nonumber
\left[1-\tanh\left(\frac{r_{\rm in}-r_{\rm c}}{\Delta r_2}\right)\right]
\frac{\eta\mu^2}{8\Omega_*r_{\rm in}^5}.
\end{eqnarray}

Taken together, (\ref{eq:sigma_ev}), (\ref{eq:visc}),
(\ref{eq:sigma_in_final}), (\ref{eq:mdot,final}) and an outer boundary
condition describe the time-dependent evolution of an accretion disc.

\subsection{Steady-State solution}
\label{sec:steadystate}

From the results of the previous sections, we can calculate the
steady-state solutions for a given \minf, the average mass
  accretion rate. For certain values of \minf, $\Delta r$ and
$\Delta r_2$, this equilibrium is unstable, leading to oscillations in
\rin~and corresponding accretion bursts.

In a steady-state, the accretion rate is constant throughout the disc,
i.e. $\dot{m}_{\rm co} = \dot{m}$:

\begin{equation}
\label{eq:rin}
\dot{m}
=\frac{1}{2}\left[1-\tanh\left(\frac{r_{\rm in}-r_{\rm c}}{\Delta
    r_2}\right)\right]\frac{\eta\mu^2}{4 \Omega r^5_{\rm in}}.
\end{equation}
Implicitly solving (\ref{eq:rin}) for $r_{\rm in}$ yields the
inner radius of the disc in a steady-state solution. 

The general steady-state surface density profile was calculated in
Section \ref{sec:ang_mom}, and is given by (\ref{eq:steady_sigma})
with an additional term since $\dot{m}\neq 0$ in the disc. The
function $f(r_{\rm in})$ is given by equation (\ref{eq:sigma0}). The
steady-state surface density profile will thus be:

\begin{eqnarray}
\label{eq:sig}
\nu\Sigma = \frac{\dot{m}}{3\pi}\left[1-\left(\frac{r_{\rm
    in}}{r}\right)^{1/2}\right] \\
\nonumber + \frac{\eta \mu^2 \Delta r}{6\pi r^4_{\rm
    in}(GMr)^{1/2}}\left[1+\tanh\left(\frac{r_{\rm in} - r_{\rm c}}{\Delta
  r}\right)\right]
\end{eqnarray}

The numerical simulations described in the following sections of the
evolution of a viscous accretion disc show that the equilibrium
solution given by (\ref{eq:rin}) and (\ref{eq:sig}) can become
unstable to episodic bursts of accretion by the process outlined in
Section \ref{sec:model}.

\subsection{Numerical setup}

To follow the time-dependent evolution of a viscous accretion disc
interacting with a magnetic field we use a 1D numerical simulation,
first making a series of mathematical transformations. 

The power-law prescription for the viscosity, (\ref{eq:visc}),
allows us to define a new function, $u$, for convenience:

\begin{equation}
u \equiv \Sigma r^{1/2+\gamma}
\end{equation}

To make our results more readily applicable to different magnetic
stars (e.g. neutron stars, magnetic white dwarves and protostars), we
adopt scale-free coordinates. The instability studied in this paper
varies on viscous time-scales of the inner disc, which are in general
much shorter than the time-scale over which the transfer of angular
momentum between the star and the disc can substantially change the
star's rotation period. A constant rotation period implies that a
constant corotation radius, making it a natural choice for scaling our
variables. We thus scale the radial coordinate to the corotation
radius, and the time in terms of the viscous time-scale ($r^2/\nu$) at
the corotation radius. Further, since we are most interested in the
behaviour of inner regions of the disc, we adopt a coordinate system
comoving with \rin:

\begin{equation}
r' \equiv \frac{r-r_{\rm in}}{r_{\rm c}}; t' \equiv t \frac{\nu_0}
{r_{\rm c}^{2-\gamma}}.
\end{equation}
Dropping the prime notation, the surface density evolution equation in
the new coordinate system then becomes:
\begin{equation}
\frac{\partial u}{\partial t} = 3 r^{\gamma -
  1/2}\frac{\partial}{\partial r}\left[r^{1/2}\frac{\partial u}{\partial
    r}\right] + \dot{r}_{\rm in}\frac{\partial u}{\partial r},
\end{equation}
with the boundary condition at \rin~given by:
\begin{equation}
u(r_{\rm in}) = \frac {\eta\mu^2}{3\pi(GM_*)^{1/2}\nu_0 r_{\rm c}^4}\frac{\Delta
  r}{r_{\rm in}}r_{\rm in}^{-3}\left[\tanh\left(\frac{r_{\rm in}-1}{\Delta
    r}\right) + 1\right].
\end{equation}
The evolution of the inner edge of the disc is given by:
\begin{eqnarray}
\label{eq:mdot}
\dot{r}_{\rm in} = \left[1-\tanh\left(\frac{r_{\rm in}-1}{\Delta
    r_2}\right)\right]\frac{\eta\mu^2}{16\pi\Omega_*\nu_0^2r_{\rm c}^{\gamma-3/2}}\frac{r^{-11/2+\gamma}_{\rm in}}{u(r_{\rm in})}\\ \nonumber
- \frac{3r^{\gamma}_{\rm in}}{u(r_{\rm in})}\frac{\partial u}{\partial
  r}\big|_{r_{\rm in}}.
\end{eqnarray}

Finally, to increase the resolution at the inner edge of the disc we
make a further coordinate transformation to an exponentially scaled
grid:

\begin{equation}
\label{eq:exp}
x \equiv
\frac{1}{a}\left[\ln\left(\frac{r-r_{\rm in}}{r_{out}-r_{\rm in}}\right) +
1\right],
\end{equation}
where $a$ is a scaling factor to set the clustering of grid points
around \rin.

We calculate the second-order discretization of the spatial
derivatives on an equally-spaced grid in $x$. To evolve the resulting
system of equations in time requires an algorithm suitable for stiff
equations. This is necessary to follow the evolution of the inner
boundary, (\ref{eq:mdot,final}). When $r_{\rm in} \gg r_{\rm c}$,
(\ref{eq:mdot,final}) reduces to a differential equation that is first
order in time. However, for $r_{\rm in} \ll r_{\rm c}$,
$\Sigma(r_{in})$ becomes very small, and the equation essentially
becomes time-independent. We have formulated the problem so that
$\Sigma(r_{\rm in})$ stays small but non-zero for all values of
$r_{\rm in}$ (so that the solutions is continuous at all values of
$r_{\rm in}$), but its small value inside $r_{\rm c}$ means that the
differential equation is stiff (since the evolution equation for
$r_{\rm in}$ in (\ref{eq:mdot,final}) contains terms of very different
sizes). To perform the time evolution, we therefore use the
semi-implicit extrapolation method (\cite{1992nrca.book.....P},
p. 724), which is second-order accurate in time and suitable for stiff
equations.

Since the grid comoves with the inner radius, the outer boundary of
our disc also moves. We set the accretion rate at the outer boundary
to be fixed in the moving coordinate system, so that it changes
slightly as the outer boundary moves. The effect is negligible as
long as the disc is large enough that the outer parts of the disc are
unaffected by the changing inner boundary condition, which we confirm
by varying the position of the outer boundary of the disc.

The solutions are sensitive to the changing conditions at the inner
boundary of the disc. To confirm that our results are robust for the
grid we have chosen, we varied the various numerical parameters of the
problem: grid resolution, the exponential stretch parameter $a$ at the
inner boundary (see (\ref{eq:exp})) and the fractional accuracy of
the solution computed by the semi-implicit extrapolation method (which sets the
maximum possible timestep).

\section{Results}
\label{sec:results}

Our primary goal in this paper is to study the conditions for which
the disc is unstable to episodic outbursts. To do this we follow the
evolution of an accretion disc in which the mean mass accretion rate,
$\dot{m}$ is a parameter of the problem by setting $\dot{m}$ as the
accretion rate through the disc's outer boundary. The other system parameters
of the problem are the stellar mass, $M_*$, frequency, $\Omega_*$, and
magnetic moment, $\mu$. The interaction between the magnetic field and
the disc introduces three additional parameters: $\eta \equiv
B_\phi/B_{\rm z}$, the fractional width of the interaction region
$\Delta r/r$, and the length scale $\Delta r_2/r$ over which the
inner edge of the disc moves from a non-accreting to accreting
state. Finally, our description of the viscosity, (\ref{eq:visc}),
introduces three additional parameters: $\alpha$, the aspect ratio of
the disc, $H/R$ (assumed constant), and $\gamma$, the radial power-law
dependence of the viscosity.

The problem has two scale invariances, which reduces the number of
free parameters. As seen in (\ref{eq:visc}), $\alpha$ and $H/R$
are degenerate. Additionally, the system parameters $\mu$, $M_*$,
$\Omega_*$ and $\dot{m}$ can be re-written as the ratio \m0mc, where
$\dot{m}_{\rm c}$ is the accretion rate in (\ref{eq:rm}) that
puts the magnetospheric radius at $r_{\rm c}$. This ratio is
equivalent to the `fastness parameter', $\Omega_{\rm in}/\Omega_*$
(where $\Omega_{\rm in}$ is the Keplerian frequency at $r_{\rm in}$)
which is sometimes used to describe disc-magnetosphere interactions.

For reference, our dimensionless parameter \m0mc~ can be
expressed in terms of physical parameters appropriate for protostellar
systems:

\begin{eqnarray}
\label{eq:refpar}
\frac{\dot{m}}{\dot{m}_{\rm c}} = \left(\frac{\dot{m}}{2.3\times10^{-7}M_\odot \rm{yr}^{-1}}\right)\left(\frac{M_*}{0.6M_\odot}\right)^{5/3}\\
\nonumber \left(\frac{B_s}{2000\rm{G}}\right)^{-2}\left(\frac{R_*}{2.1R_\odot}\right)^{-6}\left(\frac{P_*}{1~\rm{day}}\right)^{7/3}.
\end{eqnarray}

We assume that the time-averaged $B_\phi$ component will be constant
with radius in the coupled region, and set the parameter $\eta = 0.1$.
For the viscosity, $\nu = \alpha (GM_*)^{1/2}(H/R)^2 r^{\gamma}$, we take
$\alpha = 0.1$ and $H/R = 0.1$ to calculate the magnitude of $\nu_0$,
and assume $\gamma = 0.5$ everywhere in the disc. Varying $\alpha$,
$H/R$ and $\gamma$ will change the time-scale over which outbursts
occur, but will not change the general character of our outburst
solutions.

This leaves three scale-free parameters in the problem: \m0mc, $\Delta
r/r$ and $\Delta r_2/r$. We vary each of these parameters to explore
the range of unstable solutions. For small values of $\Delta r/r$ ($\sim
0.1$) and $\Delta r_2/r$ ($\sim 0.01$), and \m0mc $<$ 1, the position of
the inner boundary quickly becomes unstable and begins
oscillating. Since the position of $r_{\rm in}$ determines the mass
accretion rate on to the star, (\ref{eq:mdot_co}), the change
in $r_{\rm in}$ leads to an accretion outburst. We use the
steady-state solution (given by (\ref{eq:rin}) and (\ref{eq:sig})) as an
initial condition for all our simulations.

\begin{figure}
\includegraphics[width=\hsize]{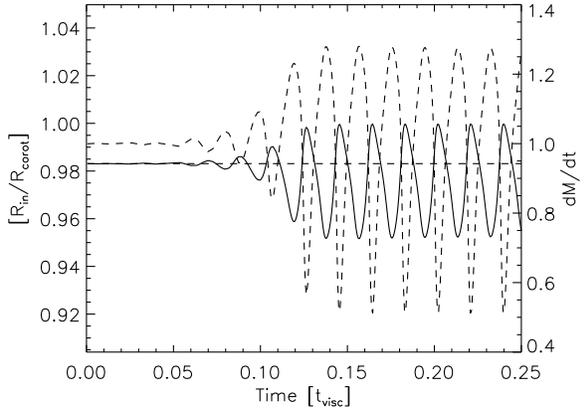}
\caption{Growth of instability from steady-state solution,
  (\ref{eq:rin}) and (\ref{eq:sig}), for \m0mc = 1, $\Delta r/r =
  0.05$, and $\Delta r_2/r = 0.014$. The inner radius (solid curve)
  evolves around its steady-state value (dashed horizontal line),
  causing the net accretion rate on to the star to change as well
  (dashed curve). \label{fig:margstab}}
\end{figure}

Fig. \ref{fig:margstab} shows the growth of the instability for
\m0mc = 1, $\Delta r/r = 0.05$ and $\Delta r_2/r = 0.014$. The solid
curve shows the evolution in $r_{\rm in}$, scaled to the corotation
radius. The horizontal dashed line shows the steady-state value for
$r_{\rm in}$. The right-hand axis plots the accretion rate on to the
star as a function of time (the dashed curve). The accretion rate is
scaled to units of the steady-state accretion rate, \minf. The
instability quickly grows out of the equilibrium solution, and
saturates into steady oscillations.

\begin{figure*}
\includegraphics{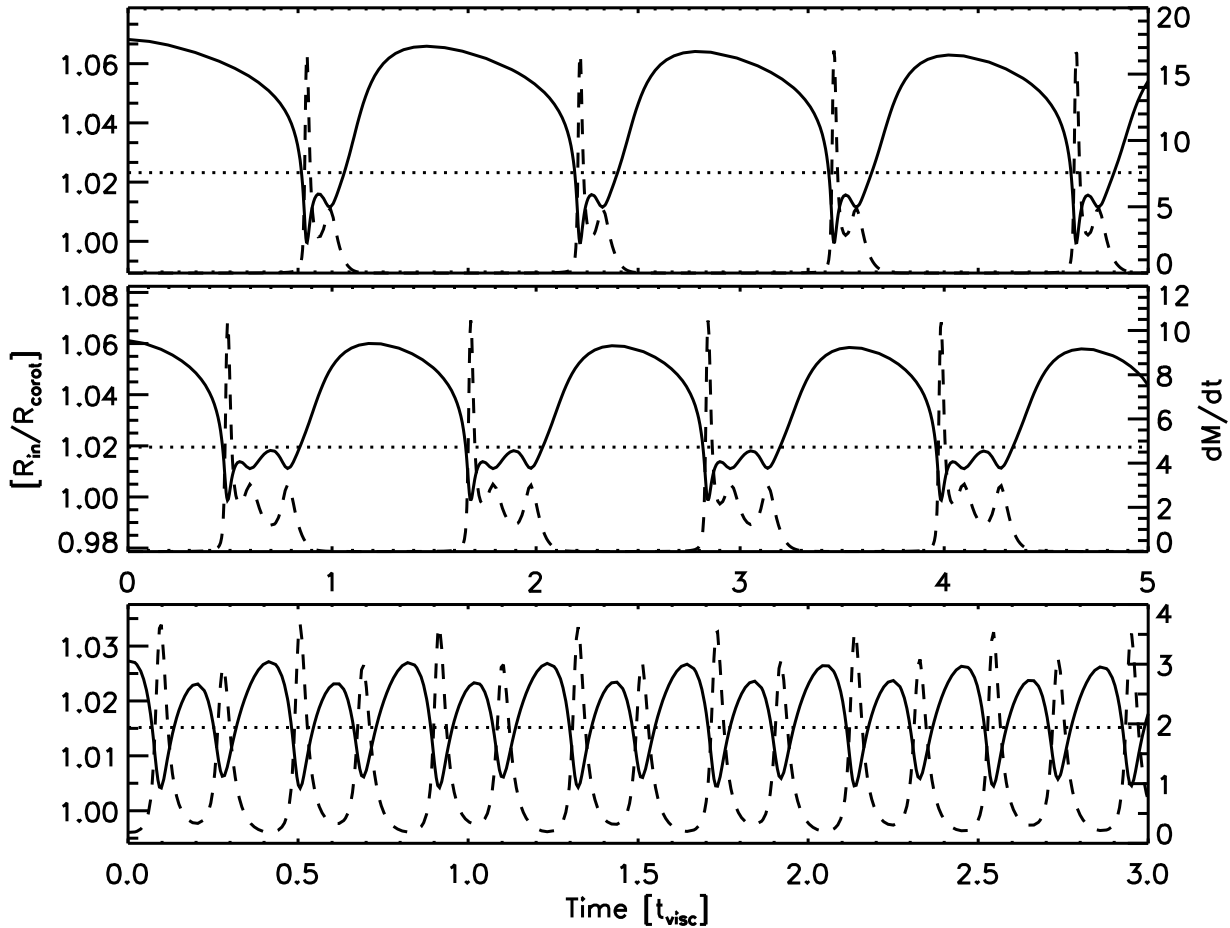}
\caption{Outburst profiles of $r_{\rm in}$ and $\dot{m}$ for moderate
  values of \m0mc. From bottom to top, \m0mc = [0.095,0.052, 0.031]. For
  adopted protostellar parameters this corresponds to $\dot{m} =
  [2.2,1.2,0.73]\times 10^{-8} M_\odot \rm{yr}^{-1}$. The lines are
  the same as in Fig. \ref{fig:margstab}.\label{fig:moderate} }
\end{figure*}

We observe a wide range of oscillatory solutions that span three
orders of magnitude in frequency, depending on the values of \m0mc,
$\Delta r/r$ and $\Delta r_2/r$. The shape of the accretion burst itself
also changes dramatically depending on the system parameters. At large
\m0mc~the bursts are quasi-sinusoidal oscillations, as in
Fig. \ref{fig:margstab} and the bottom panel of
Fig. \ref{fig:moderate}. As the mean accretion rate is decreased, the bursts
take the shape of a relaxation oscillator, where the bursts are
characterized by an initial sharp spike of accretion which then
relaxes to a quasi-steady accretion rate for the duration of the
burst, before abruptly turning off as the reservoir is emptied and
$r_{\rm in}$ quickly moves well outside $r_{\rm c}$. During the
outburst phase, higher frequency sub-oscillations are also sometimes
seen with varying intensity.

\begin{figure*}
\includegraphics{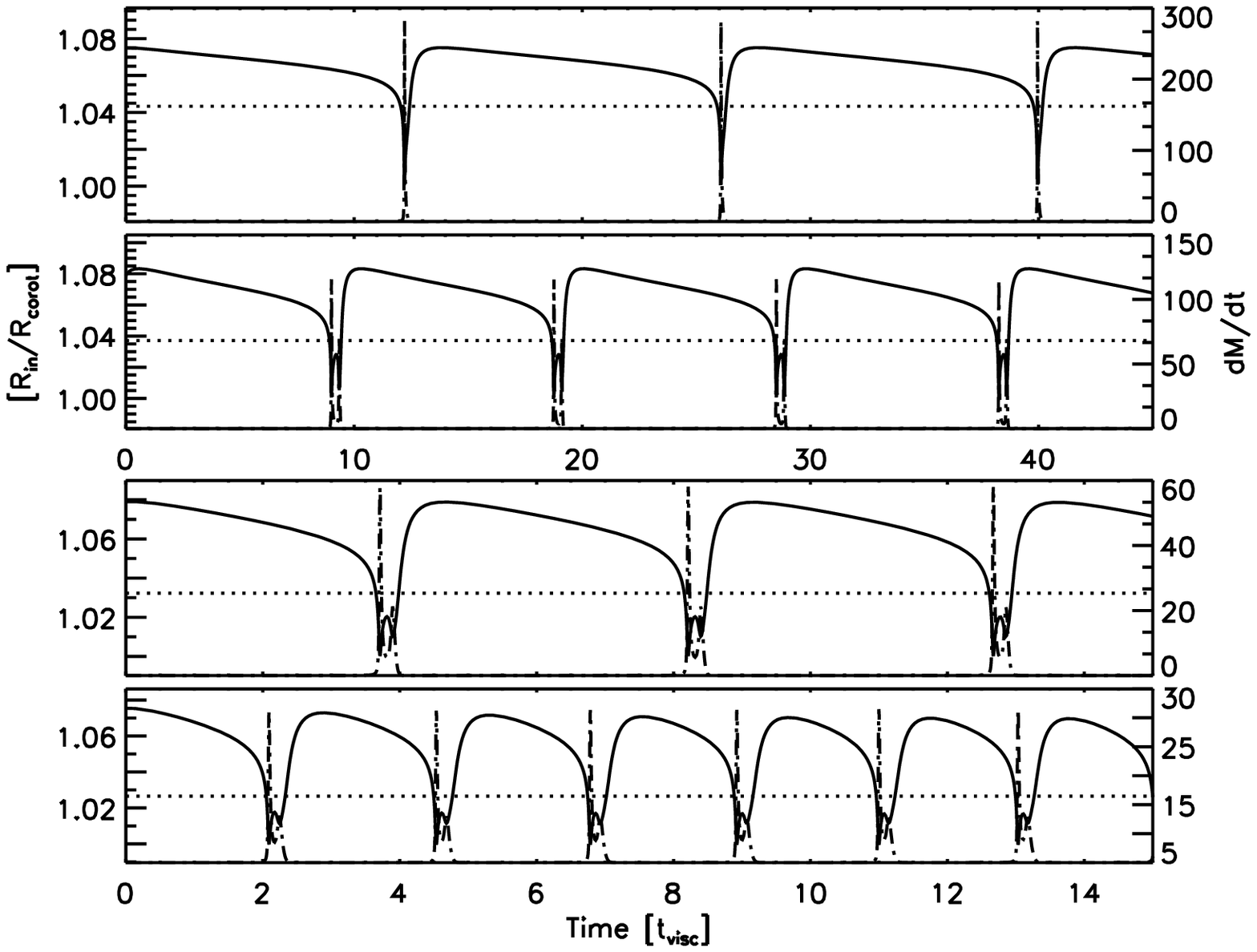}
\caption{Outburst profiles of $r_{\rm in}$ and $\dot{m}$ for
  small values of \m0mc.From bottom to top, \m0mc =
  [0.019,0.0084,0.003,0.0022]. For adopted protostellar parameters
  this corresponds to $\dot{m} = [4.5,1.9,0.95,0.38]\times
  10^{-9}M_{\odot}\rm{yr}^{-1}$. The lines are the same as in Fig.
  \ref{fig:margstab}. \label{fig:low} }
\end{figure*}

Figs. \ref{fig:moderate} and \ref{fig:low} show the evolution of
$r_{\rm in}$ and accretion rate as we vary \m0mc but the other
parameters stay fixed. From bottom to top, the panels of
Fig. \ref{fig:moderate} show the instability for \m0mc =
[0.095, 0.052, 0.031] ($\dot{m} = [2.2,1.2,0.73]\times 10^{-8} M_\odot
\rm{yr}^{-1}$ for the parameters in (\ref{eq:refpar})). At the
highest mean accretion rate, $r_{\rm in}$ (the solid curve) oscillates with
a high frequency around its steady-state value (dashed line), with
corresponding bursts of accretion on to the star (dashed curve). As
\m0mc~is decreased, the accretion profile changes to much lower
frequency outbursts, with long periods of quiescence as $r_{\rm in}$
moves away from $r_{\rm c}$ and accretion ceases completely. The
high-frequency oscillation that dominates for \m0mc = 0.095 is
superimposed over the low-frequency accretion bursts for lower
\m0mc. Fig. \ref{fig:low} shows the continuation of
Fig. \ref{fig:moderate} for \m0mc~ = [0.019,0.0084,0.003,0.0022]
($\dot{m} = [4.5,1.9,0.95,0.38]\times 10^{-9}M_{\odot}\rm{yr}^{-1}$).
The characteristic accretion burst profile essentially stays the same
as \m0mc~is decreased, with sharp spikes at the beginning and end of an
accretion outburst. The overall amplitude of the outburst decreases
only slightly with decreasing mean accretion rate. The initial spike
decreases by about 20\% as the mean accretion rate drops from \m0mc = 0.052
to \m0mc = 0.0022. The more significant effect is that the length of
time between outbursts increases with decreasing \m0mc, since at low
average accretion rates it takes longer to build enough mass to drive
another outburst. The overall shape of the outburst is relatively
insensitive to changing \m0mc, becoming shorter as \m0mc~
decreases. At the lowest accretion rate ($3.8\times 10^{-10}
M_\odot\rm{yr}^{-1}$; the top panel of Fig. \ref{fig:low}), the burst
consists of only one sharp spike. As we have formulated the problem,
the instability will persist down to arbitrarily low accretion rates.

\begin{figure*}
\includegraphics{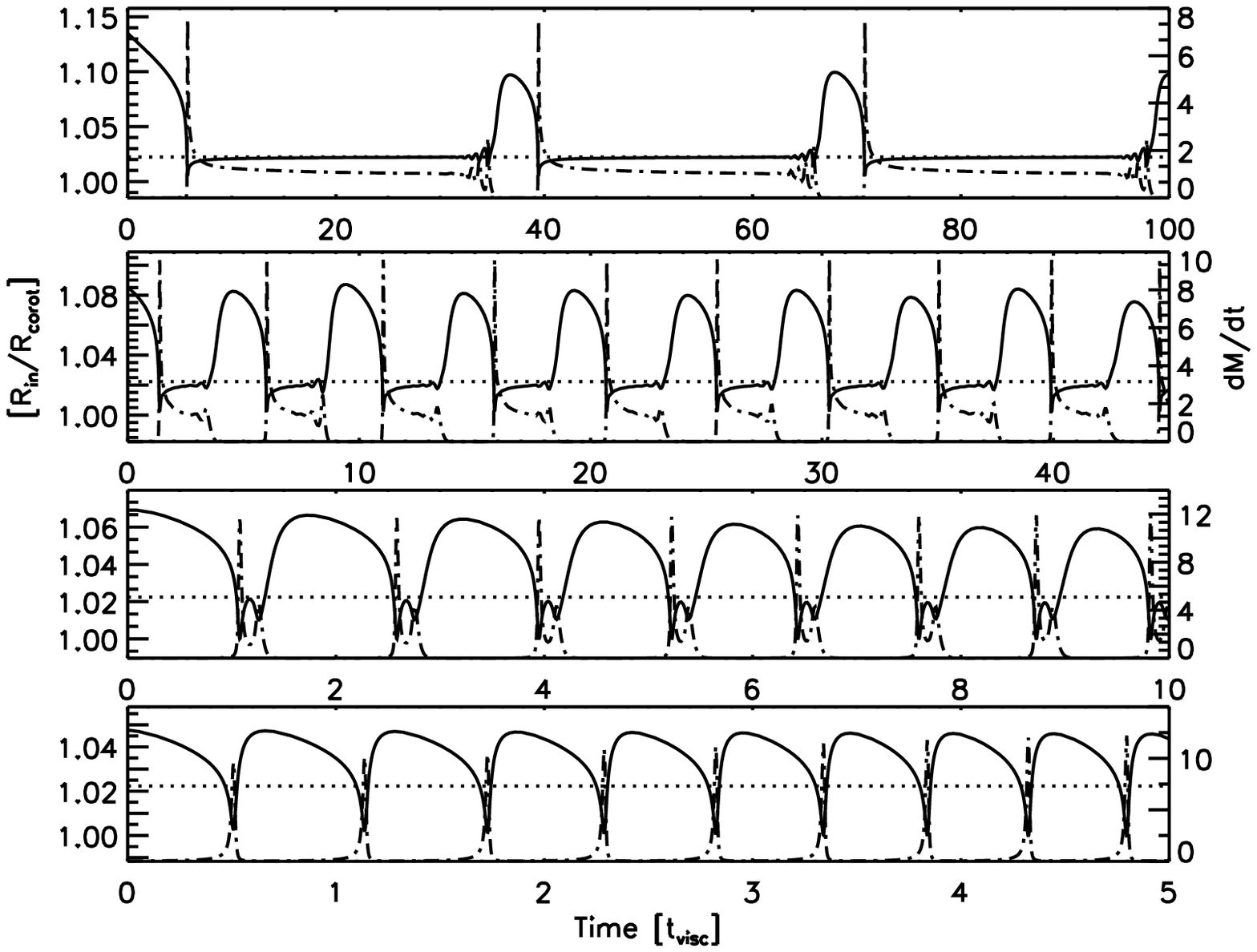}
\caption{Outburst profiles of $r_{\rm in}$ and $\dot{m}$ for changing
  $\Delta r/r$, with $\Delta r_2/r = 0.014$ and \m0mc = 0.04. From bottom
  to top, $\Delta r/r = [0.03,0.05,0.07,0.09]$. The lines are the same
  as in Fig. \ref{fig:margstab}.\label{fig:deltar}}
\end{figure*}

Changing the other parameters, $\Delta r/r$ and $\Delta r_2/r$, has a
much stronger effect on the shape of the outburst than changing the
mean accretion rate. Fig. \ref{fig:deltar} shows the outburst profiles
for different values for $\Delta r/r$, setting \m0mc = 0.04 and
$\Delta r_2/r = 0.014$. From the bottom to top, $\Delta r/r$ =
[0.03,0.05,0.07,0.09], which spans the unstable region of $\Delta r/r$
for the adopted \m0mc. For small $\Delta r/r$ the instability
manifests itself as repeating short bursts of accretion, with
comparatively long quiescent phases. As $\Delta r/r$ increases, the
frequency of the outburst decreases, and the duty cycle increases
dramatically. For very large $\Delta r/r$ the outburst lasts about 200
times as long as for the minimum $\Delta r/r$ but at lower accretion
rate after the initial spike.  The burst profile of the instability is
thus sensitive to small changes in $\Delta r/r$, but the range in
$\Delta r/r$ over which the instability exists is quite small.We find
a similar range of outburst profiles by changing $\Delta r_2/r$ and
keeping $\Delta r/r$ fixed, except with the opposite trend: for large
$\Delta r_2/r$ the instability manifests as a series of short spiky
bursts, becoming longer as $\Delta r_2/r$ decreases.

We next considered the parameter space in \m0mc, $\Delta r/r$ and
$\Delta r_2/r$ over which the instability occurs. We have briefly
explored the effect of varying both $\Delta r/r$ and $\Delta r_2/r$
over a small range in \m0mc~ and found that, although the outburst
profile changes somewhat, the range over which $\Delta r/r$ and
$\Delta r_2/r$ produce unstable solutions are independent. We
therefore assume that $\Delta r/r$ and $\Delta r_2/r$ vary
independently of each other for all \m0mc, and consider the range of
the instability over the [\m0mc, $\Delta r/r$] and [\m0mc, $\Delta
  r_2/r$] spaces separately.

\begin{figure}
\includegraphics[width=\hsize]{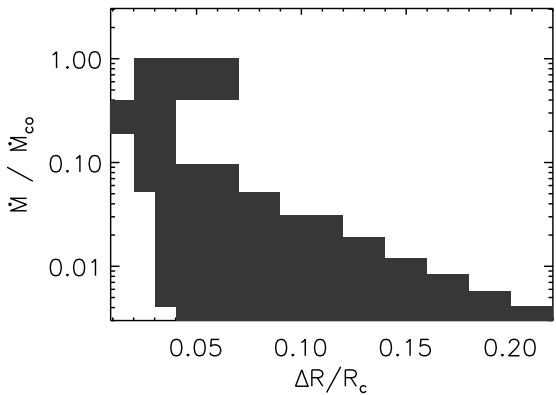}
\caption{Parameter map of instability as a function of \m0mc and
  width of interaction region $\Delta r/r$, with constant $\Delta r_2 =
  0.014$. The shaded regions denote unstable
  parameters. \label{fig:parmap1}}
\end{figure}

Fig. \ref{fig:parmap1} shows the range of unstable solutions (shown
as shaded regions)  changing \m0mc~and $\Delta r/r$, but keeping $\Delta
r_2/r $ fixed at 0.014. Although there is a small unstable branch around
\m0mc~= 1, in general as $\Delta r/r$ increases, a lower \m0mc~is
required before the instability sets in. 
\begin{figure}
\includegraphics[width=\hsize]{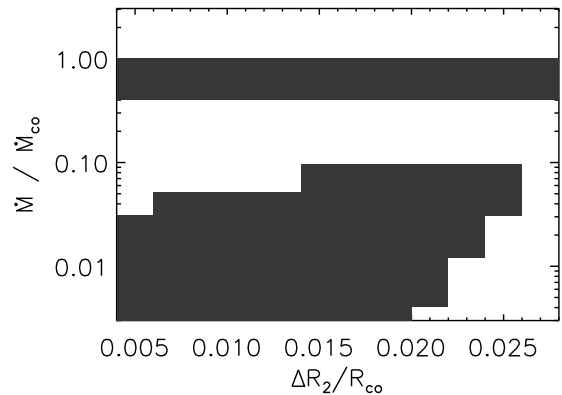}
\caption{Parameter map of instability as a function of \m0mc~and
  accretion transition length $\Delta r_2/r$, with constant $\Delta r_2
  = 0.014$. The shaded regions denote unstable
  parameters. \label{fig:parmap2}}
\end{figure}

Fig. \ref{fig:parmap2} shows the unstable solutions changing
\m0mc~and $\Delta r_2/r$ but keeping $\Delta r/r$ fixed at 0.05. The
opposite trend from Fig. \ref{fig:parmap1} is seen, with a larger
range of unstable accretion rates. There is again a range of unstable
solutions around \m0mc = 1, although in this case the unstable region
extends over the entire $\Delta r_2/r$ parameter space. The instability
likely extends to smaller $\Delta r_2/r$, but we do not explore the
region smaller than $\Delta r_2 = 0.005$ on physical grounds, since
such a small transition length will likely be unstable to other
instabilities like the interchange instability (see Section
\ref{sec:constraints}). As with changing $\Delta r/r$, the outburst
profile changes substantially over the small range of $\Delta r_2/r$ in
which the instability occurs.

\section{Discussion}
\label{sec:discussion}

In this paper we studied a disc instability first explored by
\cite{1977PAZh....3..262S} and ST93, with a more physically motivated
and general formulation of the problem than was used in ST93. In
particular, we have improved the description of the disc-field
interaction when the disc is truncated outside corotation by deriving
conditions for a `quiescent' state, in which the angular momentum
transferred from the star into the disc halts accretion altogether. In
agreement with ST93, we observe a wide range of oscillatory behaviour,
and the frequency range of individual outbursts spans three orders of
magnitude.

The period of the cycle seen in
Figs. \ref{fig:margstab}--\ref{fig:low} varies from 0.02 to 20$t_{\rm
  c}$, where $t_{\rm c}$ is the nominal viscous time-scale at the
corotation radius $t_{\rm c}=r_{\rm c}^2/\nu(r_{\rm c})$. Though cycle
times scale with $t_{\rm c}$, this is evidently not the only
factor. As discussed in ST93, the viscous time-scale relevant for the
cycle period depends on the size of the disc region involved. This
depends itself on the cycle period, hence the period must be
determined by additional factors. One of these is the mean accretion
rate, but the physical conditions in the magnetosphere-disc
interaction region have an equally important effect.

From Figs. \ref{fig:parmap1} and \ref{fig:parmap2} it appears that
there are two different kinds of instability. One of these operates in
a narrow range of accretion rates, around the value where steady
accretion would put the inner edge at corotation. The instability in
this case is of the type shown in Fig. \ref{fig:margstab}: an
approximately sinusoidal modulation, characteristic for a weak form of
instability. The inner edge of the disc oscillates about a mean value,
but stays inside the width of the transition region. The longer cycles
in the upper parts of Figs. \ref{fig:parmap1} and \ref{fig:parmap2} are
a strongly non-linear, relaxation type of oscillation. The inner edge
is somewhat outside the transition region for much of the cycle with
no accretion taking place (the `quiescent' phase), and dips in for a
brief episode of accretion before moving back out again. This is the
kind of cycle envisaged by \cite{1977PAZh....3..262S}. During the
quiescent phase, the disc (\cite{1977PAZh....3..262S} call it a
`dead disc') extracts angular momentum from the star by the magnetic
interaction at its inner edge. These two forms of instability are
merged into a continuum in ST93, as a result of the different (and
less realistic) assumptions made there about the interaction between
disc and magnetosphere outside corotation. This difference also
affects the dependence on the mean accretion rate. Whereas in ST93
cyclic behavior was found only in a limited range of accretion rates,
our results show that cycles can occur in principle at arbitrarily low
accretion rates, with steadily increasing cycle period and decreasing
duty cycle of the accretion phase.

Figs. \ref{fig:moderate} and \ref{fig:low} show that the radius of the
inner edge of the disc does not move by more than 10\% around
corotation, even at the lowest mean accretion rates. For example in
the case $\dot m/\dot m_{\rm c}=9.5 \times 10^{-2}$ of
Fig. \ref{fig:moderate}, the standard `ram pressure' estimate would
yield a much larger magnetosphere radius, about $r_{\rm m}=3.6\,r_{\rm
  c}$. The difference arises because in our cyclic accretion states
the conditions in the inner disc are very different from those assumed
in conventional estimates of $r_{\rm m}$; the density in the inner
disc, for example, is much higher.

At $r_{\rm in}\le 1.1 r_{\rm c}$, the velocity difference between the
magnetosphere and the disc is only 5\%, much less than the 40\% which
mass would need in order to escape from the system. `Propellering' of
mass out of the system is thus unlikely to be effective. This does not
exclude that some mass loss (powered by a magnetic wind from the disc
or the interaction region around the inner edge of the disc) may also
take place, but our results show that this is not a necessary
consequence for a disc in what is traditionally called `propeller'
regime.

At sufficiently low accretions rates one would expect, however, that
propellering would also be a possible outcome: if the rotation rate of
the star is high enough, matter could be ejected before it has the
time to form a dense disc. The existence of a cyclic form of accretion
at low accretion rates thus suggests that two different accretion
states are possible, and that there would be a second parameter
determining which of the two is realised. This might simply be the
history of the system.

If a disc is initially absent and accretion is started, the density
will initially be low enough that ejection by propellering can prevent
accretion altogether. The cataclysmic variable AE Aqr (e.g.
\citealt{1997MNRAS.286..436W}) is likely to be such a case. On the
other hand, if a disc is initially in a high accretion state such that
the inner edge is inside corotation, a subsequent decline to low
accretion rates could lead to the cyclic accretion described
here. Such a situation could be at work in the TTauri star EX Lupi
(where the initial high accretion phase has ended). It could also be
appropriate for the X-ray millisecond pulsar, SAX J1808.8-3658, which
has shown a 1-Hz QPO in the decline phase of several outbursts
\cite{2009ApJ...707.1296P}.  The pile-up of mass at the magnetosphere
will maintain the disc this state, and prevent propellering even when
the mean accretion rate drops to very low values.

The instability studied in this work has not yet been observed in
numerical simulations, partly because most numerical simulations do
not run for long enough to observe it, but mainly because most
simulations have focused on either accreting or strong propeller
cases. However, in virtually all numerical simulations outflows and
variability in the disc are observed, with an intensity that varies
between different simulations. Gas pile-up at the inner edge of the
disc is also observed, with the amount of pile-up tied to the
effective diffusivity of magnetic field at the inner edge of the disc
(e.g. \cite{2004ApJ...616L.151R}). The process of closing and opening
field lines provides a source of mass to launch both a
weakly-collimated outflow (the disc wind) and a well-collimated jet
(e.g \citealt{1996ApJ...468L..37H, 1997ApJ...489..199G,
  2009arXiv0907.3394R}). The whole cycle takes place on time-scales
that can vary between the dynamical and viscous time-scales at the
inner edge of the disc, but are generally of higher frequency than the
disc instability studied in this paper. The inner edge of the disc
also oscillates significantly (although it remains on average outside
corotation), from between a few stellar radii
\citep{2009arXiv0907.3394R} up to 30 stellar radii
\citep{1997ApJ...489..199G}. Even if such variability is present, the
instability studied in this paper can still occur provided the
outflows/accretion bursts generated by field lines opening are not
strong enough to fully empty the reservoir of matter accumulating just
outside $r_{\rm c}$.

\section{Conclusions}
\label{sec:conclusion}

We have studied the accretion of a thin viscous disc on to a
magnetosphere of a magnetic star, under the influence of the magnetic
torque it exerts on the disc. We focused in particular on cases with
low accretion rates. For high accretion rates such that the inner edge
$r_{\rm in}$ of the disc is inside the corotation radius, standard
steady thin viscous disc solutions are recovered. However, when the
inner edge is near corotation we find that the accretion becomes
time-dependent, and takes the form of cycles consisting of alternating
accreting and non-accreting (`quiescent') states. The period of this
cycle varies from a small fraction of the characteristic viscous time
scale in the inner disc, $r_{\rm in}^2/\nu$, to a large multiple of it,
depending on the mean accretion rate as well as on the precise
conditions assumed at the magnetosphere.

These cyclic accretion solutions continue to exist indefinitely with
decreasing accretion rate. The cycle period increases, while the duty
cycle of the accreting phase decreases with decreasing accretion
rate. In the quiescent phase after a burst of accretion, the inner
edge of the disc moves outward, and mass starts piling up in the inner
regions of the disc. In response, the inner edge eventually starts
moving back in again and accretion picks up as $r_{\rm in}$ crosses
the corotation radius. This empties the inner regions of the disc,
causing the inner edge to move outward again. The cycle thus has the
properties of a relaxation oscillator, as found before in ST93. The
reservoir involved is the mass in the inner region of the disc.  These
results (as well as those of \cite{1977PAZh....3..262S} and ST93) show
that accretion without mass ejection can occur at accretion rates well
inside what is usually called the `propeller' regime. Instead of the
mass being ejected, the accreting mass can stay piled up at high
surface density in the inner disc, just outside corotation.  We have
suggested that systems with very low accretion rates can be in either
of these states. Propellering would occur when a disc is initially
absent and mass transfer is first initiated (the case of AE Aqr for
example), while a system with an accretion rate that drops from an
initially high value would end in the cyclic accretion state described
in this paper. This would apply to most cataclysmic variables and
X-ray binaries, as well as some TTauri stars.

\section{Acknowledgments}

CD'A would like to thank Stuart Sim for useful scientific discussion,
and acknowledges financial support from the National Science and
Engineering Research Council of Canada.

\bibliographystyle{mn2e}
\bibliography{magbib}
\label{lastpage}
\end{document}